\documentclass[fleqn,10pt]{wlscirep}
\usepackage[utf8]{inputenc}
\usepackage[T1]{fontenc}
\usepackage{subfig}
\usepackage{textcomp}

\title{Predicting Fault Slip via Transfer Learning}

\author[1,2]{Kun Wang}
\author[1]{Christopher W. Johnson}
\author[1]{Kane C. Bennett}
\author[1,*]{Paul A. Johnson}
\affil[1]{Geophysics Group, Earth and Environmental Sciences Division, Los Alamos National Laboratory, Los Alamos, NM 87545 USA}
\affil[2]{Center for Nonlinear Studies, Los Alamos National Laboratory, Los Alamos, NM, 87545, USA}

\affil[*]{paj@lanl.gov}


\begin{abstract}
Data-driven machine-learning for predicting instantaneous and future fault-slip in laboratory experiments has recently progressed markedly due to large training data sets. In Earth however, earthquake interevent times range from 10's-100's of years and geophysical data typically exist for only a portion of an earthquake cycle. Sparse data presents a serious challenge to training machine learning models. Here we describe a transfer learning approach using numerical simulations to train a convolutional encoder-decoder that predicts fault-slip behavior in laboratory experiments. The model learns a mapping between acoustic emission histories and fault-slip from numerical simulations, and generalizes to produce accurate results using laboratory data. Notably slip-predictions markedly improve using the simulation-data trained-model and training the latent space using a portion of a single laboratory earthquake-cycle. The transfer learning results elucidate the potential of using models trained on numerical simulations and fine-tuned with small geophysical data sets for potential applications to faults in Earth.
\end{abstract}

\begin{document}

\flushbottom
\maketitle

\thispagestyle{empty}

\section*{Introduction}

In Earth, predicting instantaneous and future characteristics of fault slip has long been a fundamental goal of geoscientists from an earthquake hazards perspective, but also to improve the basic understanding of fault mechanics \cite{scholz_2019}. Recent progress towards these goals has been achieved by applying a variety of machine learning (ML) approaches \cite{bergen2019machine, ren2020machine} in the laboratory using shear experimental data to describe physical properties \cite{rouet2017machine, rouet2018estimating, lubbers2018earthquake, hulbert2019similarity, jasperson2019unsupervised, bolton2019characterizing, zhou2019earthquake} and in the Earth using geophysical data to characterize episodic slow-slip that occurs in subduction zones \cite{rouet2020probing, hulbert2019similarity} as well as transform faults \cite{johnson2020learning}. In shear experiments, earthquakes or ``labquakes'', generated during a single experiment produce a sufficiently large data set for training ML models. However, on natural faults the repeat cycles for all but the smallest earthquakes can span timescales on the order of decades to hundreds of years. Thus, \emph{in-situ} geophysical measurements as input for data-driven ML models are generally not available or sufficiently complete for more than a portion of a single earthquake cycle. In particular this problem exists for large magnitude (M>7) earthquakes that produce strong, damaging ground motions. This conundrum presents a serious challenge if the goal is to use data-driven modeling techniques to characterize the physics of fault slip throughout the complete earthquake cycle and to advance earthquake hazards assessment. 

A type of model generalization known as transfer learning \cite{pan2009survey, goodfellow2016deep} is one potential solution to overcome the problem of data sparsity. Generalizing ML models using transfer learning has been explored in a number of areas in geophysics; for instance in seismology applications, transfer learning has been used to improve nonlinear and ill-posed inverse problems associated with seismic imaging or subsurface feature classification \cite{chevitarese2018transfer, siahkoohi2019importance, cunha2020seismic, zhang2020data}. To our knowledge, no attempt has been made to apply transfer learning using data from numerical simulations to make quantitative predictions of fault slip in laboratory experiments or Earth observations. 

In this work we develop a deep learning convolutional encoder-decoder (CED) model that employs a time-frequency representation of acoustic emissions (AE) from numerical simulations and laboratory shearing experiments. We describe results from the CED model transfer learning and show the successful application of this technique for multiple data sets. 

The long term goal is to develop a ML approach that characterizes seismogenic faults. In Earth we generally record only a portion of a slip cycle on a fault and one could apply a similar approach of transfer learning and cross-training. If such a procedure works at the laboratory scale, it may be applicable to Earth, despite the dramatic changes in scale and complexity of seismogenic faults.

\section*{Results}

\subsubsection*{Transfer learning from numerical simulations to laboratory shear experiments}

\begin{figure}[htbp]
\centering
\includegraphics[width=0.85\linewidth]{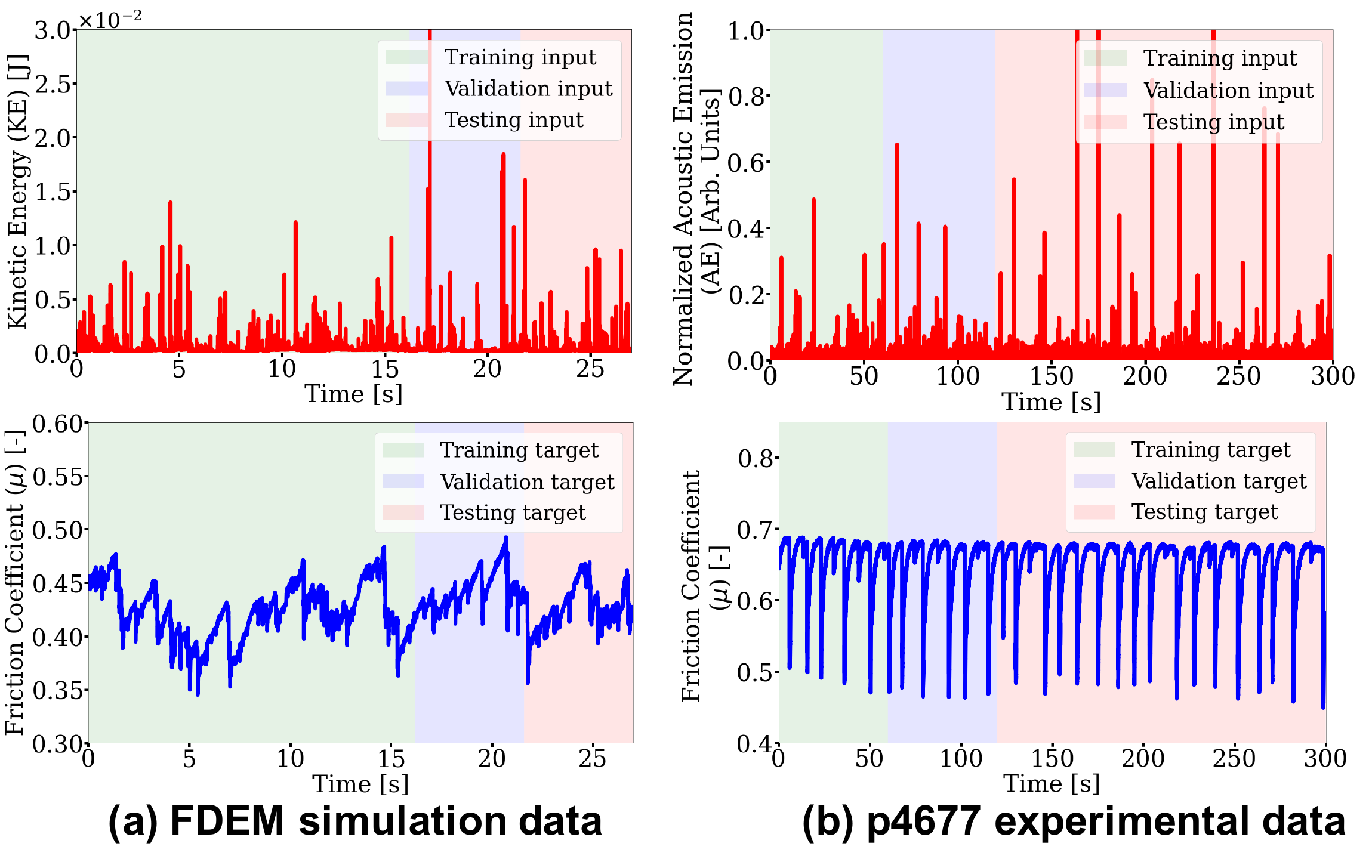}
\caption{Numerical simulation and experimental data used in the transfer learning analysis. The top row shows the deep learning model input signal as the kinetic energy and acoustic emissions, respectively, and the bottom row shows the target friction coefficient.   (a) FDEM time series are split into training/validation/testing segments (60/20/20\%) shown in green, blue, and pink shades, respectively. The convolutional encoder-decoder is fully trained and tested using these data. (b) The experimental data (p4677) are split into training/validation/testing segments (20/20/60\%) to include 6 cycles of stick-slip events for training the model latent space.}
\label{fig:simulationVSexperiment}
\end{figure}

The laboratory data \cite{johnson2013acoustic, bolton2019characterizing} is from a bi-axial shear device that consists of a slider-block bounded by fault gouge and external blocks through which a confining load is applied. A constant shear velocity is applied and when the system approaches steady state conditions, repetitive stick-slip motion occurs (see Figure \ref{fig:simulationVSexperiment}b). The bi-axial device simultaneously measures acoustic emission (AE) and the normal and shear stresses required to calculate the bulk friction coefficient. 

The numerical simulation \cite{gao2018modeling} applies a combined Finite-Discrete Element Method (FDEM) model of a fault-shear apparatus resembling the bi-axial device used in the laboratory experiments described here \cite{geller}. The input training data to the CED model from simulation is the kinetic energy, which is a proxy for the measured continuous AE signal in the bi-axial shear experiment. Changes in seismic moment are reflected in variations observed in the kinetic energy; therefore, the kinetic energy represents the kinematic behavior of the granular fault system. The CED label data is the bulk friction coefficient between the sliding blocks. 

In summary, the data from these sources, numerical simulation and laboratory experiment (Figure \ref{fig:simulationVSexperiment}), are used by the CED model. The supervised learning method is a regression procedure, using the AE from experiment or the kinetic energy from simulation to predict the instantaneous characteristics of slip, specifically the coefficient of friction.

\begin{figure}[htbp]
\centering
\includegraphics[width=0.9\linewidth]{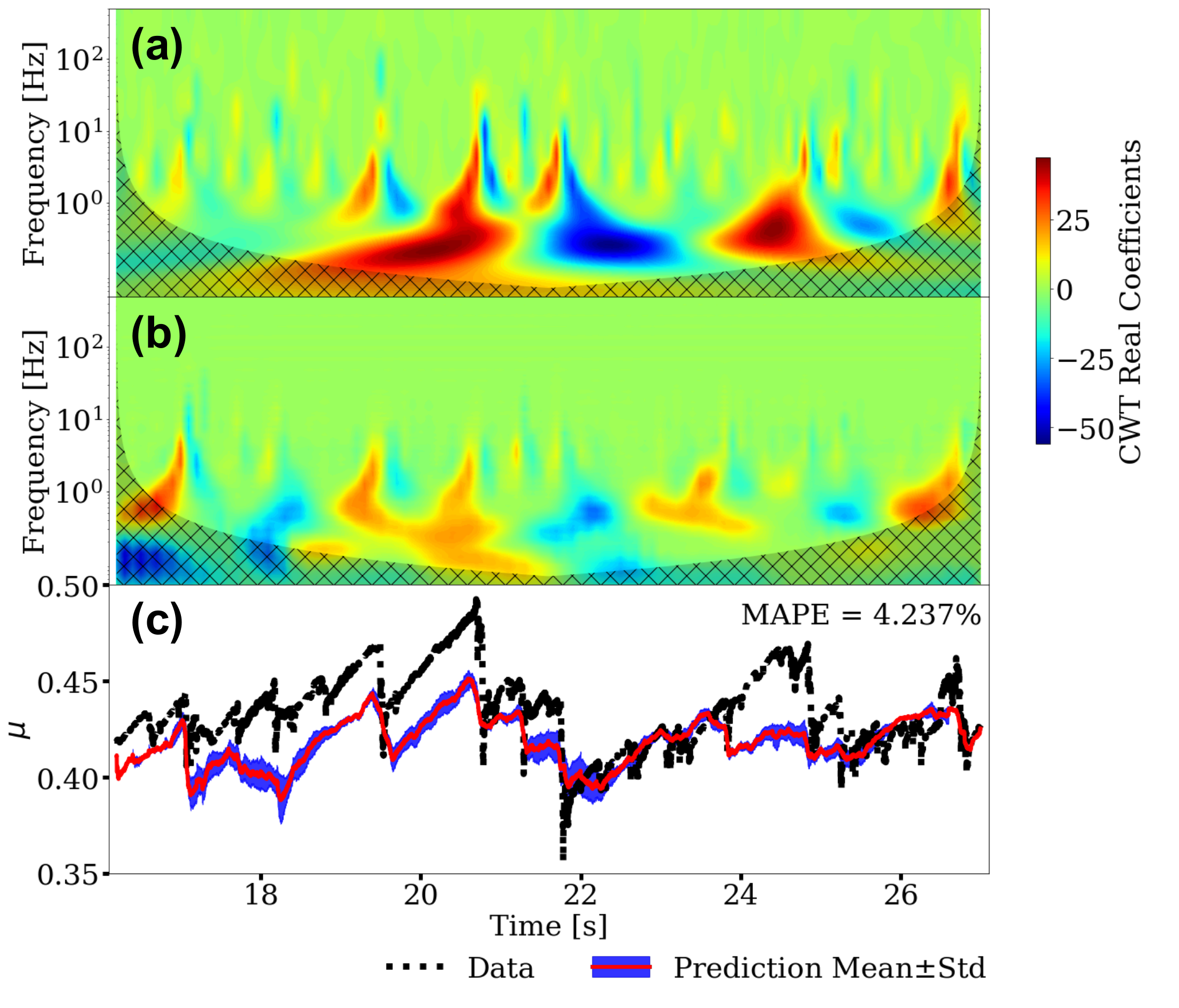}
\caption{Instantaneous frictional coefficient prediction from the CED model trained on FDEM simulation data. The (a) input and (b) predicted scalograms are shown. The cross-hatched region in (b) indicates the cone of influence where edge effects are important. The predictions from the CED are made applying sliding windows with 2 s length and step size of 0.2 s. The predicted scalogram is the average of all sliding windows. (c) Ground truth versus prediction: the numerical simulation data (black line) and model-predicted friction coefficient from the inverse of the scalogram (red line) is shown with the blue region indicating 1-standard deviation for the predictions in the overlapping windows. The Mean Absolute Percentage Error (MAPE) is listed for the numerical simulation and predicted values.}
\label{fig:ced_model_mu_predict_fdem}
\end{figure}

As a point of reference for the transfer learning approach, the results shown in Figure \ref{fig:ced_model_mu_predict_fdem} are produced by training, validating, and testing entirely on the numerical simulation data. The scalogram represents the average performance among all trained models. The predicted friction coefficient captures the general slip trends including many frictional failures. However, the results are modest as reported by the Mean Absolute Percentage Error (MAPE) of 4.237\% for the numerical simulation data. 

\begin{figure}[htbp]
\centering
\includegraphics[width=0.9\linewidth]{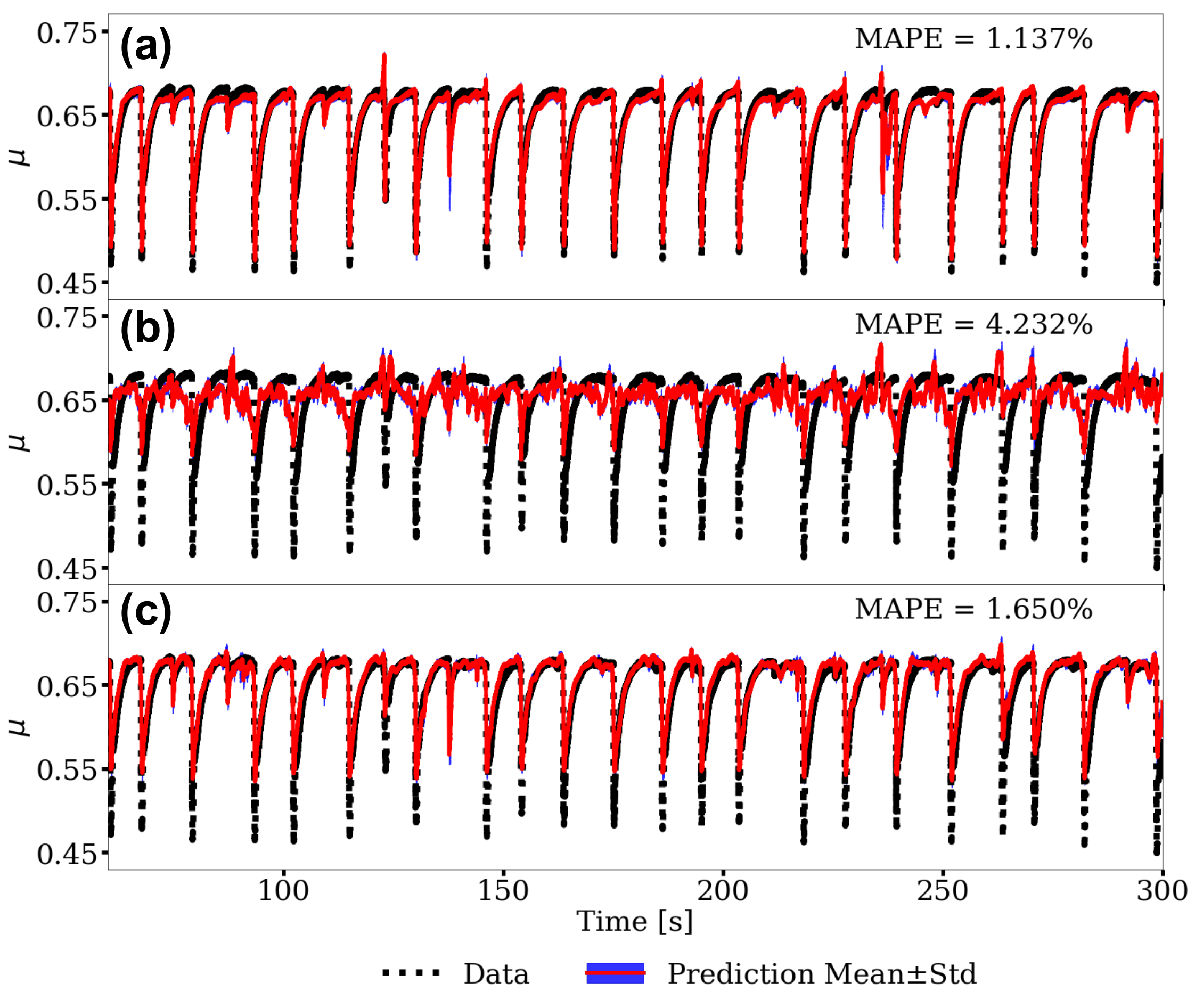}
\caption{Predictions from CED models.  (a) Model  trained and tested on the laboratory data.  (b) Model trained on simulations and tested on laboratory data. The first 20\% of the AE signal (0 s to 60 s) was used for training to construct the model. (c) Cross-trained model.  Model trained on the simulations, then fixing the encoder and decoder layers, with the model then additionally trained on the bottleneck (latent space) applying a portion of the laboratory (experiment p4677) data.}
\label{fig:transfer_model_mu_predict_p4677}
\end{figure}

The same procedure is followed using only the laboratory AE and friction data to train a separate CED model. The first 20\% of the AE signal (0-60 s) is used for training data. The friction predictions from the testing data produce a MAPE of 1.137\% (Figure \ref{fig:transfer_model_mu_predict_p4677}a). The model performs very well estimating the variations in friction coefficients predicting the frictional failures associated with slip events. 

For the first transfer learning exercise, we use the model trained on simulation and apply it to predict the friction in the laboratory experiment. The trained model uses the experiment AE as input and the label is the friction from experiment. We emphasize that the CED model never sees experimental data during training with the simulation data.  The predictions show a decrease in performance with a MAPE of 4.232\% (Figure \ref{fig:transfer_model_mu_predict_p4677}b) when compared to the model trained solely on the laboratory data. The maximum friction drop, which has an equivalence to event moment (moment = GAu; where G is the gouge shear modulus, A is slip area and u is the fault displacement), is consistently less than that measured from the experiment. Under-prediction of the event moment is a common problem with many ML models when applied to the bi-axial shear data \cite{johnson2020learning}, without considering transfer learning. Nonetheless, we find the timing and scale of the predictions are surprisingly good considering the significant differences between the FDEM numerical simulation and the laboratory shear experiment.

With an eye to faults in Earth where obtaining sufficient training data is a challenge, we introduce transfer learning by cross-training the model. This approach is an extension of an established  transfer learning technique used in image classification tasks, e.g., \cite{yosinski2014transferable, shin2016deep} (As applied to image classification, the convolutional layers of a model are pre-trained on a large database (e.g., ImageNet \cite{deng2009imagenet, huh2016makes} and then specific convolutional layers are extracted and held constant, then merged with an additional fully-connected classification layer that is trained with data specific to the problem of interest.). The resulting predictions shown in Figure \ref{fig:transfer_model_mu_predict_p4677}c and are very good with MAPE=1.650\%, which is a significant improvement from the MAPE of 4.232\% before cross-training. The model predictions are now comparable to the MAPE of 1.137\% obtained when training directly on p4677 laboratory data.

\begin{figure}[htbp]
\centering
\includegraphics[width=0.9\linewidth]{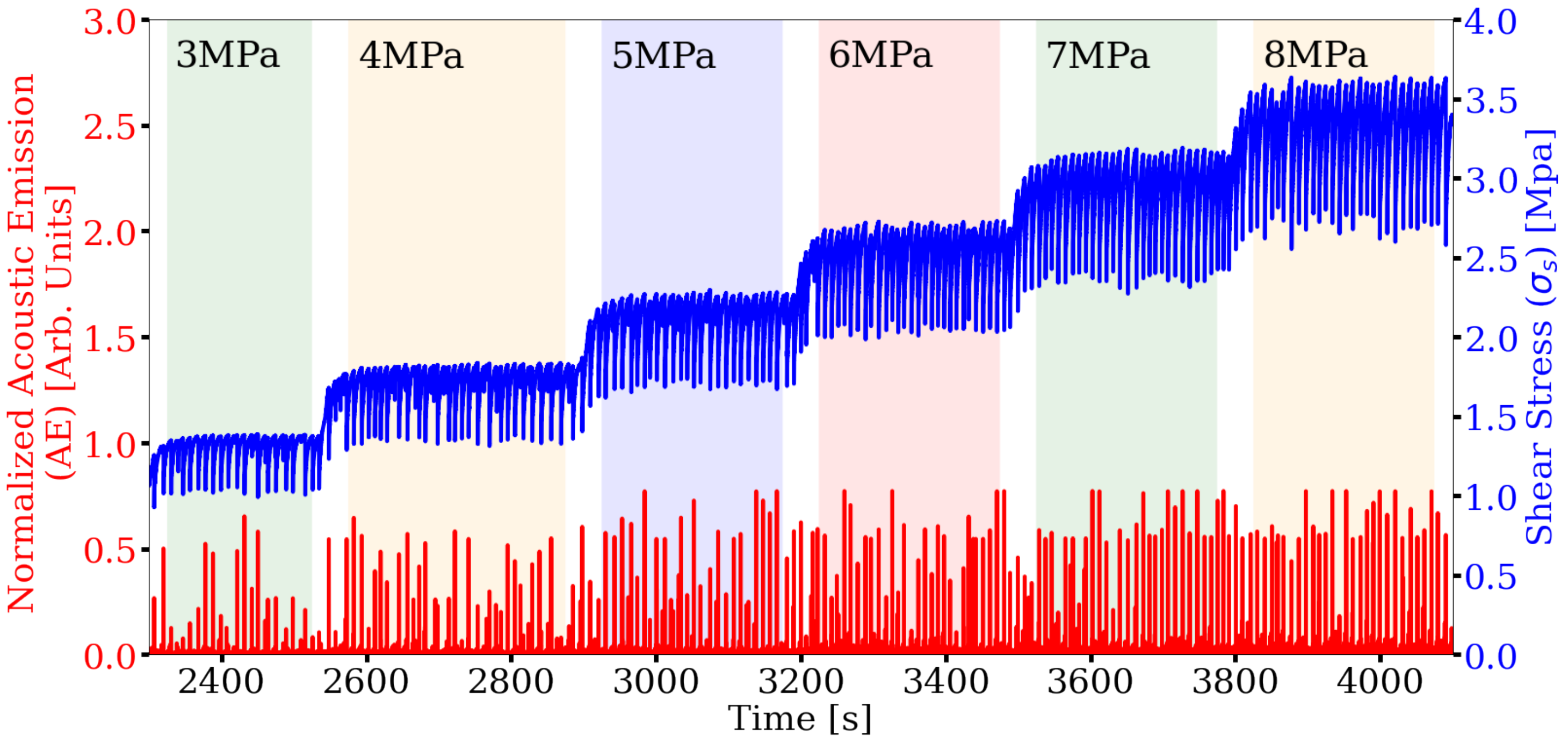}
\caption{CED Model generalization. The cross-trained model is rigorously validated using a second laboratory experiment (p4581) as an independent data set. The input signal is the acoustic emission (AE, in red) and the target signal is the friction derived from the shear stress ($\sigma_s$, in blue) at progressively increasing applied normal loads (3-8 MPa), shown in sequence and delineated by different shading. Model predictions are shown in Figure \ref{fig:transfer_model_mu_predict_p4581}.}
\label{fig:p4581_data}
\end{figure}

As a more rigorous test on how well the cross-trained CED model predicts the laboratory experiments, we apply the identical model to a different laboratory experiment conducted in the bi-axial apparatus. These experimental data are never seen during training of the latent space. This second experiment was conducted over a range of confining loads (normal stress) from 3-8 MPa (Figure \ref{fig:p4581_data}). The only information applied from the different confining load levels are the mean and standard deviation statistics used to normalize the AE and $\mu$ signals when producing the input scalograms to the model and when reconstructing the output scalograms. 

\begin{figure}[htbp]
\centering
\includegraphics[width=0.9\linewidth]{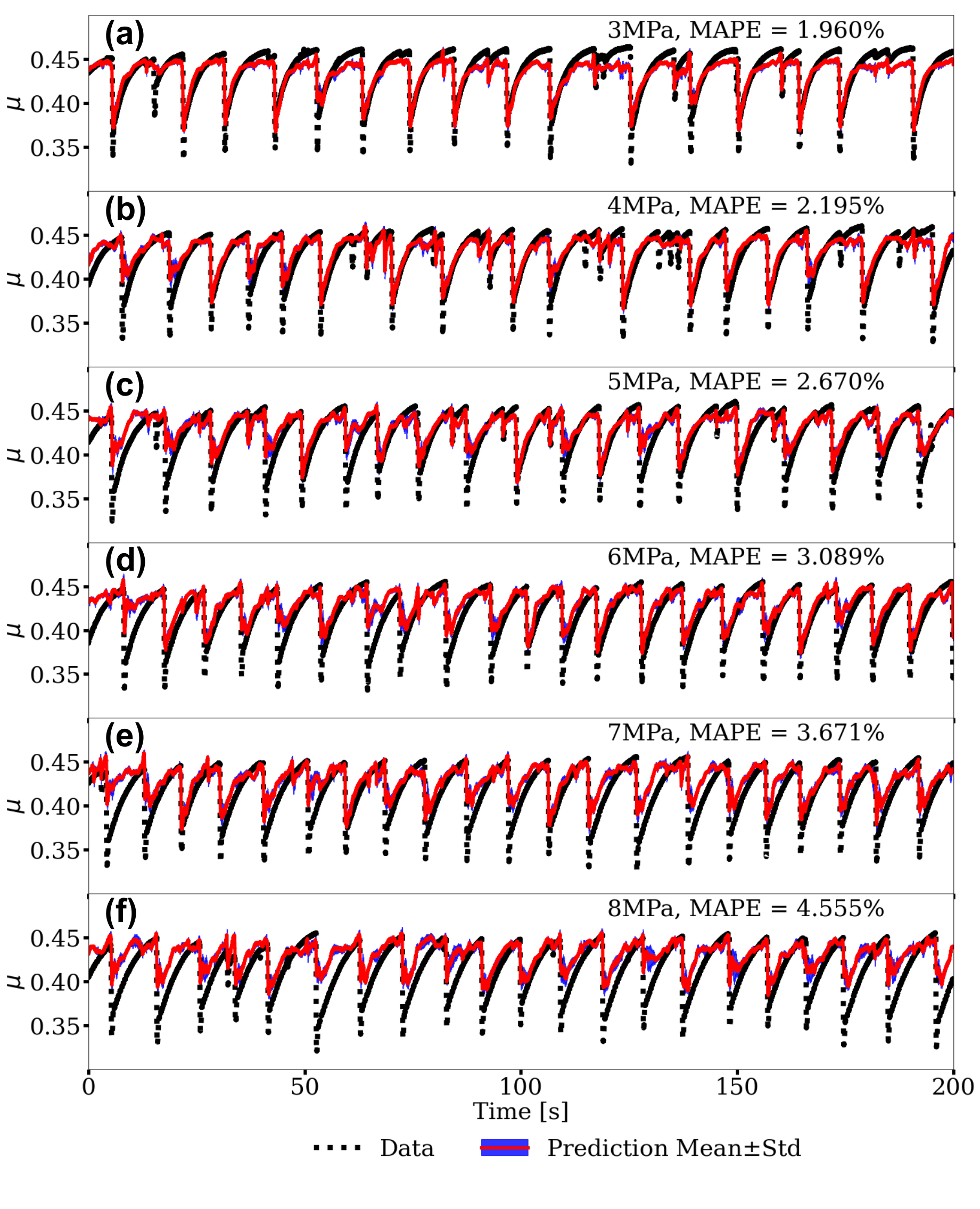}
\caption{Transfer learning applied to an independent experiment. Shown are predictions from the cross-trained CED model (experiment p4581) with normal loads that progressively increase from 3MPa to 8MPa (see Figure \ref{fig:p4581_data}). Each load level is predicted independently using the cross-trained model from simulation (the encoder and decoder) and data from experiment (p4677) conducted at 2.5 MPa.
The predictions as manifest by the MAPE progressively decrease with increasing load level. Nevertheless, the results show that the transfer learning approach with cross-training of the latent space, which accounts for only 20\% of the total CED model parameters, is a powerful approach to predicting the frictional state of the experimental fault.}
\label{fig:transfer_model_mu_predict_p4581}
\end{figure}

The predictions applying the cross-trained model to the second experiment are shown in (Figure \ref{fig:transfer_model_mu_predict_p4581}). The results are remarkably good as indicated by the MAPE's.  The 3 MPa data exhibits the best MAPE, presumably because the confining load is close to the 2.5 MPa value in the p4677 experiment that was used to train the latent space (Figure \ref{fig:transfer_model_mu_predict_p4581}a). The model predictions as manifest by the MAPE increase with increasing normal loads. The prediction errors appear to be due primarily to the poor predictions of the frictional failure magnitudes. Nonetheless, the instantaneous slip-event times are captured at all load levels, as are the stress buildups during inter-event periods. 

\subsubsection*{Transfer learning with extremely limited data in laboratory experiments}
\begin{figure}[htbp]
\centering
\includegraphics[width=0.9\linewidth]{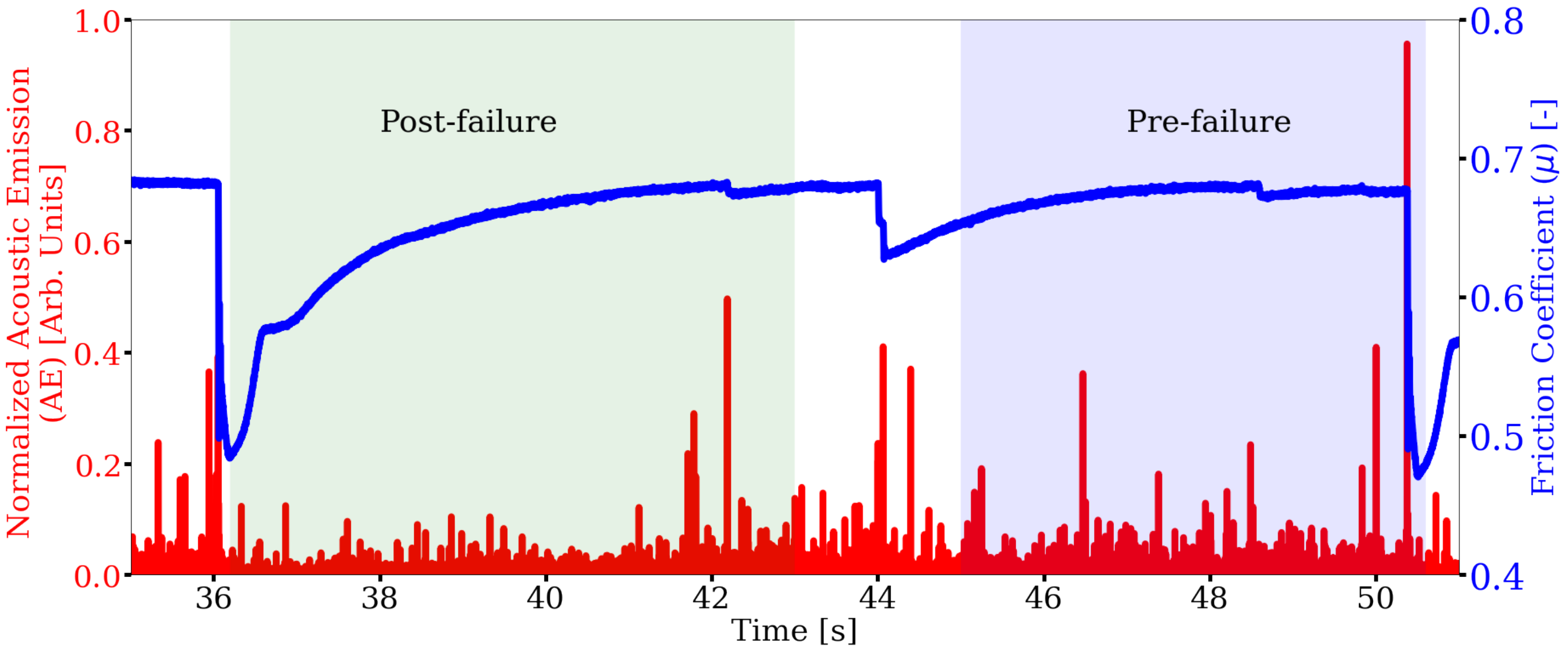}
\caption{Testing the limits of the transfer learning:  CED model with cross-training of the latent space applying limited portions of AE data from a single slip cycle. The model input signal is acoustic emission (AE), and the output signal is the friction coefficient ($\mu$). The green region shows the \emph{post-failure} data used for the transfer learning exercise, and the blue region shows the \emph{pre-failure} data (see Figure \ref{fig:transfer_model_mu_predict_onecycle_train} for model results).}
\label{fig:transfer_model_p4677_onecycle_data}
\end{figure}

Because slip cycles in Earth are so long (decades to 100's of years) and we rarely have more than a portion of associated seismicity within a full seismic cycle, we conduct a cross training exercise that mimics this data-poor circumstance. We do so by using only limited portions of a single slip cycle from the laboratory experiment for training the model latent space. Specifically, we train the latent space by applying only the \emph{post-failure} or the \emph{pre-failure} $\mu$ signals from experiment p4677 data (Figure \ref{fig:transfer_model_p4677_onecycle_data}). The post-failure comprises the time-period when the shear stress is increasing relatively rapidly following the previous slip event. The pre-failure period comprises the period when the fault is late in the cycle, near-critical state, and beginning to nucleate \cite{johnson2013acoustic}. The model encoder and decoder trained with the simulation data again remain unchanged. The model is trained and validated using 90\% and 10\% of the data, respectively, for the pre-failure and post-failure analysis. Because the available data only spans a short time interval, the size of the sliding windows is reduced from 2s to 0.4s and the step size is reduced from 0.2s to 0.1s. The training is terminated when the validation does not reduce for 100 epochs to prevent over-fitting. 

\begin{figure}[htbp]
\centering
\includegraphics[width=0.75\linewidth]{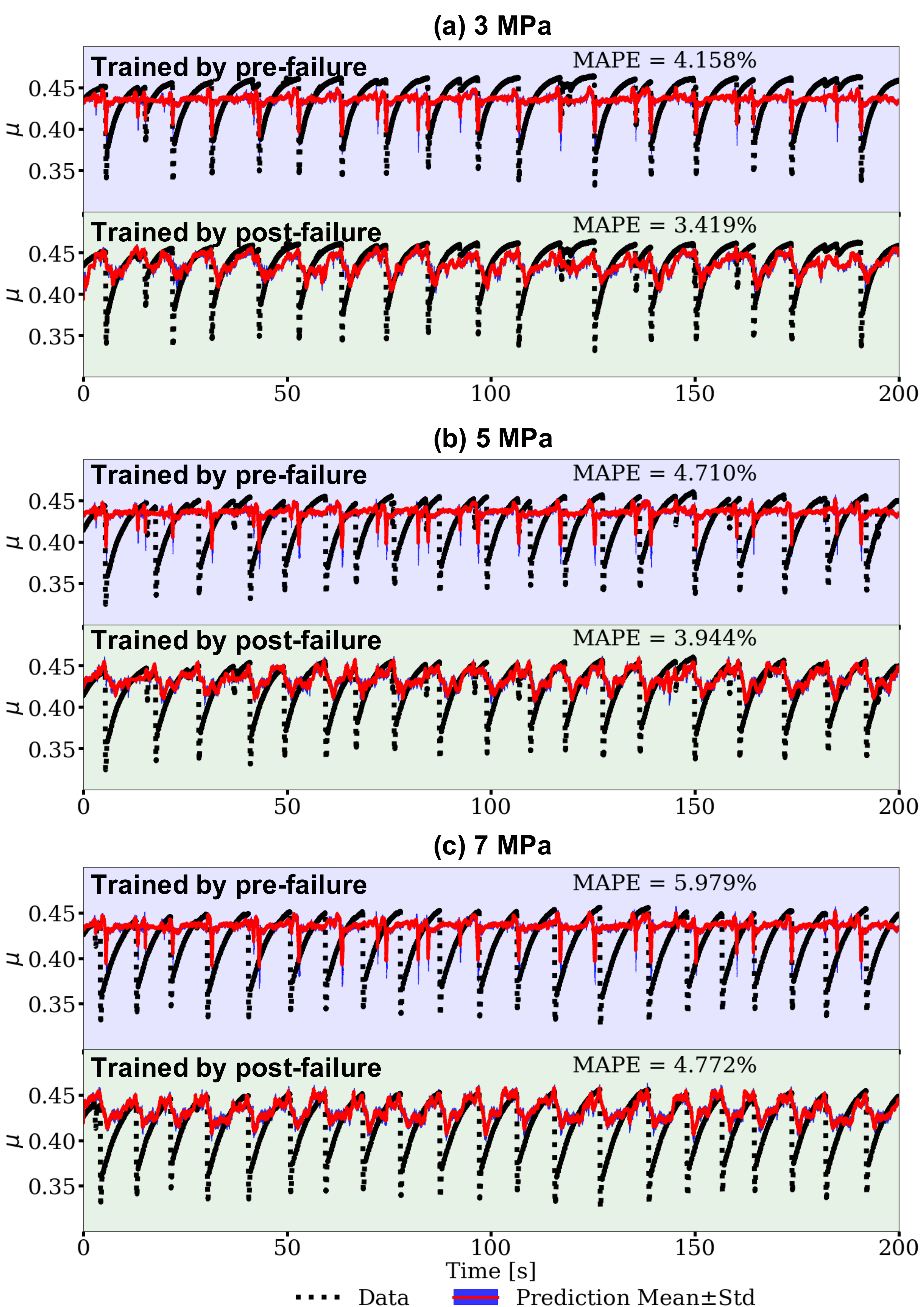}
\caption{Model cross-training applying limited portions of the experimental slip cycle. Two models are trained. In both cases only the latent space is trained using data from experiment p4677 applying the following (see Figure \ref{fig:transfer_model_p4677_onecycle_data}. One model is trained applying AE data from the `\emph{post-failure} portion of the slip cycle, comprising the time-period when the shear stress is increasing relatively rapidly. The second model is trained applying AE data from the \emph{pre-failure} period comprising the period the fault is in a critical state and nucleating. The model encoder and decoder trained applying the simulation data remain unchanged. Friction predictions on data from the experiment p4581 testing data set for pre-failure at applied loads of 3,5, and 7 are shown in each top row of (a, b, c) and post-failure in each bottom row of (a, b, c). The color in the panels correspond to the colored training data in Figure \ref{fig:transfer_model_p4677_onecycle_data}.}
\label{fig:transfer_model_mu_predict_onecycle_train}
\end{figure}

After cross-training the latent space using the two data sets from experiment p4677 (post-failure and pre-failure) to produce two separate CED models, the models are used to predict the friction in the second experiment, p4581, for 3MPa, 5MPa and 7MPa applied load, on the post-failure and pre-failure signals. The results are shown in Figure \ref{fig:transfer_model_mu_predict_onecycle_train}.  As before, the magnitude of the frictional failures are not well predicted---otherwise the trained models perform surprisingly well in both cases. The result applying the post-failure data is slightly better than that from the pre-failure training data. This suggests the model has learned more frictional states during latent-space training. As anticipated, the model using experiment p4677 and trained applying 6 full cycles provides the most robust result; however, the model results with extremely limited training cross-training data are very encouraging.

\subsubsection*{Transfer learning predicting time to failure (TTF) in laboratory experiments}

\begin{figure}[htbp]
\centering
\includegraphics[width=0.9\linewidth]{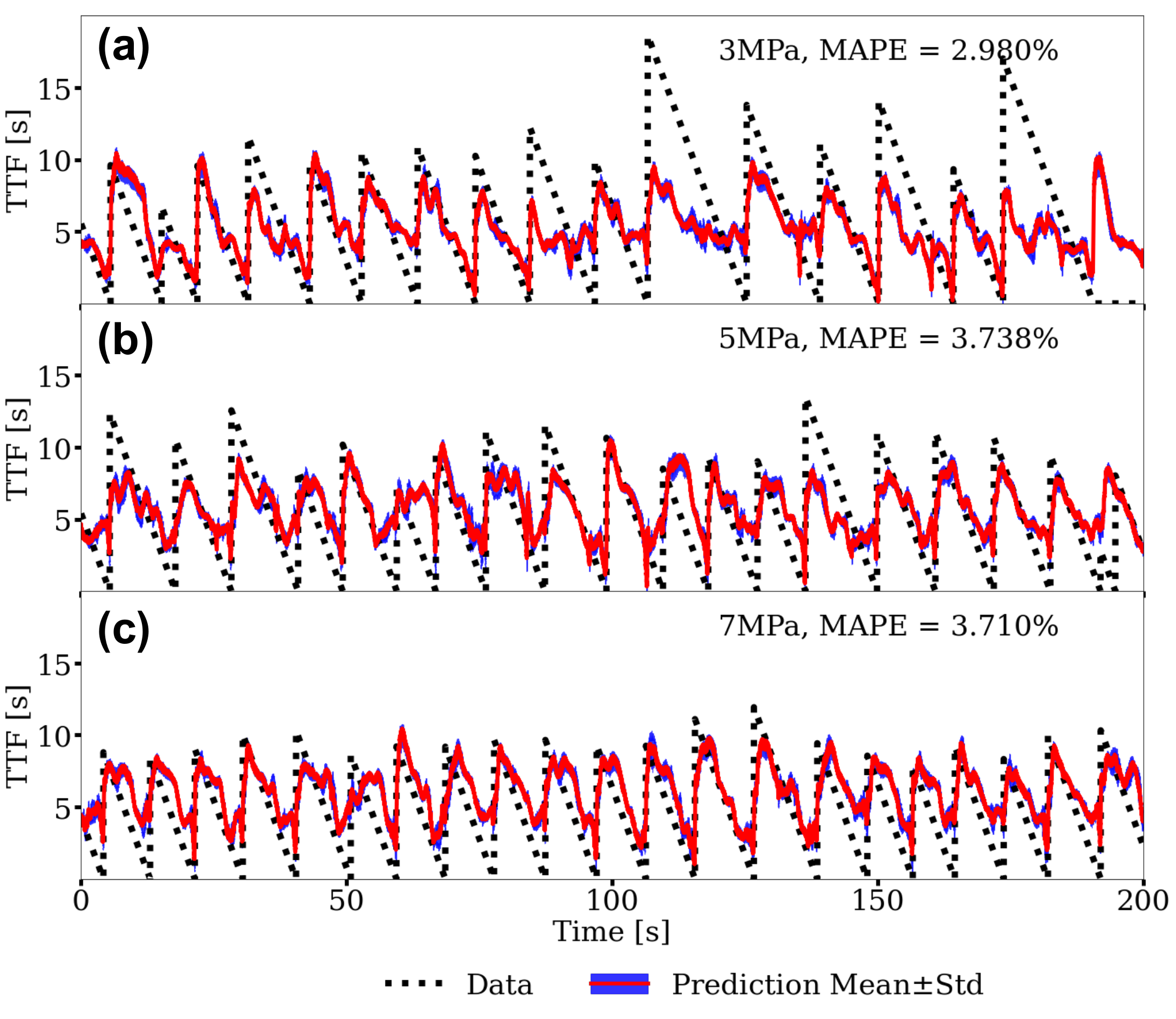}
\caption{Predictions of time-to-failure (TTF) applying the transfer learning, cross-trained model. Predictions of TTF for laboratory p4581 data at (a) 3 MPa, (b) 5 MPa, and (c) 7 MPa confining loads using transfer learning of the CED model trained on FDEM data. Only the latent space is trained on laboratory p4677 data for the TTF predictions.  The blue dashed line shows the ground truth TTF derived from the experiment. The red curve shows the model predicted TTF.}
\label{fig:transfer_model_t2f_predict}
\end{figure}

Transfer learning can also be applied to predicting other output time series. Here we showcase the predictions on the signals of time-to-failure (TTF). Failure times are defined as when the time derivative of the $\mu$ signal is below -10 $/s$. The raw AE signal is used as input, just as for the instantaneous predictions of the friction coefficient. Data from p4581 are again used for testing purposes only. The predictions are illustrated in Figure \ref{fig:transfer_model_t2f_predict} for 3MPa, 5MPa and 7MPa load levels. The predictions are good, if not perfect considering the task, as underscored by their respective MAPE scores. Indeed they are notable considering the they are obtained from cross-training and transfer learning. 

\section*{Discussion}
The predictions of the instantaneous friction obtained applying transfer learning from FDEM simulations to laboratory data from the bi-axial shear device are surprisingly good. When model cross-training is then applied, the predictions improve significantly. Further, when we apply the same cross-trained model to the second experiment conducted at multiple applied loads, the model predictions are still surprisingly good---there exists a larger misfit with increasing load, but the timing of the event is accurate regardless of the under-prediction in friction failure magnitude. The results are even more remarkable considering the FDEM simulation was not meant to directly simulate the experiment---material properties and dimensions of the fault gouge and shear-blocks were considerably different. Indeed, the results indicate the simulations contain an AE (kinectic energy) evolution captured in the spectral characteristics that can predict the actual AE in experiment. The results suggest the simulations, despite the differences, provide a sufficient range of behaviors for the models to learn and reproduce the laboratory behavior---the FDEM simulation exhibits more complex slip behaviors than the experiments in regards to the range of interevent times. Consequently, the trained model is able to predict the simpler behavior with more quasi-periodic interevent times exhibited by the experimental data. The slip frictional failure magnitude predictions are less accurate than the timing---the full range of frictional failures observed during sliding is under-predicted. Knowing this, one could conduct simulations that produce larger frictional failures to determine if this improves the laboratory failure predictions. 

To summarize the significant highlights, when the model latent space is additionally trained applying long segments of the experimental AE data, the predictions improve markedly. As the latent space training data set is decreased to a portion of a single cycle, the predictions are still reliable, even when the trained model is applied to a different experiment and at varying applied load levels. Further, the time-to-failure predictions applying the cross-trained model are also surprisingly good.  

Overall, the results are very promising and suggest further improvements are possible. For real-world seismic applications, the stick-slip repeat cycles can be on timescales ranging from several decades to centuries, and generally the available field recordings only cover a partial slip event. We imagine a scenario in Earth in which we have a temporally limited set of observations used in tandem with fault simulations to train a similar type of model. This may be of great value as we address evolving fault slip and earthquake hazards in the real Earth. 

\section*{Methods}
\subsection*{Numerical simulation and laboratory experiments} 
Numerical simulations of a laboratory experiment performed by Gao et al. \cite{gao2018modeling} were obtained by applying the combined finite-discrete element method (FDEM) using the Hybrid Optimization Software Suite package (HOSS) \cite{knight2020hoss}.  
The FDEM was originally developed by Munjiza \cite{Munjiza} to simulate continuum to discontinuum transitional material behavior. FDEM combines the algorithmic advantages of the discrete element method with those of the finite element method. In FDEM, each discrete element is comprised of a subset of finite elements that are allowed to deform according to the applied load, which is particularly useful in capturing deformations in the fault gouge material as well as at the gouge particle--plate boundary. 

The FDEM model was applied to simulate a two-dimensional, photoelastic shear laboratory experiment conducted by Geller and others \cite{geller}. Two-dimensional plane stress conditions were assumed and the model comprised 2,817 circular particles confined between two identical plates. Three thousand bi-dimensional particles with diameters of 1.2 and 1.6 mm were used, respectively (1,500 of each). The plates had dimensions of 570 $\times$ 250 mm. At the plate interfaces semi-circular shaped `teeth' were placed to increase friction between plates. The particles had Young's modulus and Poisson's coefficient of 0.4 GPa and 0.4, respectively, while the plates had Young's modulus and Poisson's coefficient of 2.5 MPa and 0.49, respectively, far smaller than those used in the bi-axial shear experiment described below. Shearing velocity was 0.5 mm/s.

The laboratory data \cite{dieterich1984effect, marone1998laboratory, niemeijer2010frictional, johnson2013acoustic, scuderi2016precursory} were obtained from a bi-axial shear apparatus. 
Laboratory experiments fail in quasi-periodic cycles of stick and slip that mimic to some degree the seismic cycle of loading and failure on tectonic faults. The apparatus comprises a central steel block that is driven at fixed loading velocity of 10 {\textmu}m/s for the experiment. This loading imparts shear stresses within two gouge layers that are 100 mm square with an initial thickness of 5 mm. The gouge layers are located on either side of the central driving block and confined by a second steel layer of 20 mm thickness. The gouge consists of monodisperse glass beads of 104–149 {\textmu}m diameter with Young's modulus of 70 GPa and Poisson coefficient of 0.3; the steel blocks have Young's modulus of approximately 180 GPa and Poisson's coefficient of approximately 0.29.  A load‐feedback servo control system maintains a fixed normal stress of 2.5 MPa for experiment p4677, while measuring shear stress throughout the experiment. For experiment p4581, progressively larger loads were applied, and at each load level, steady state was achieved before a change to the successive load level. The shearing speed was 5 mm/s for both experiments. Mechanical data measured on the apparatus throughout the experiments included the shearing block velocity, the applied load, the gouge layer thickness, the shear stress, and the coefficient of friction. Continuous AE emissions from the fault zone seismic wave radiation were recorded with piezo-ceramics embedded inside blocks of the shear assembly \cite{riviere2018evolution}.

We note that in the FDEM simulations considered here the AEs were not propagated in the model. We assume the kinetic energy obtained from the fault simulations as being equivalent to the AEs recorded on the experimental shear apparatus based on previous analysis \cite{gao2018modeling}. Used here as an equivalent quantity to the AE is the kinetic energy ($E_k$) summed from the entire system. Since the plates and particles work together as an ensemble, it is the aggregate energy evolution that governs the stick-slip behavior in granular fault gouge. In the bi-axial experiment, the source of the AE signal is at the grain contact level \cite{trugman2020}. Fault gouge contacts broadcast AE independently and/or simultaneously \cite{trugman2020}, and displace the sideblocks equivalently along the dimensions of the block due to the extreme stiffness of the steel, in analogy to the $E_k$ behavior in the simulation. Thus, the $E_k$ is approximately equivalent to the magnitude of the continuous AE time series (norm of the acoustic emission recorded by the two channels of the lab experiments), which is the source of elastic waves. We say approximately because there is a very modest amount of wave dissipation during wave propagation in the experiment from the fault gouge layer through the steel plates to the detectors.

\subsection*{Training, validation and testing data} 
The continuous time series signals (AE, kinetic energy, and friction coefficient) from the experiment and FDEM simulation are converted into scalograms. 

The Continuous Wavelet Transform (CWT; see reference \cite{torrence1998practical} for a comprehensive description of the method) procedure is applied to the laboratory experiment p4677 acoustic emission (AE) and friction ($\mu$) time series to produce training/validation/testing data. 

Scalograms are calculated for the laboratory experiment p4581 and only used as testing data for experiments conducted at different normal stresses.

\section*{Acknowledgements}
KW, KCB and PAJ acknowledge support by the U.S. Department of Energy, Office of Science, Office of Basic Energy Sciences, Chemical Sciences, Geosciences, and Biosciences Division under grant 89233218CNA000001. KW acknowledges support by the Center for Nonlinear Studies (CNLS) at Los Alamos National Laboratory. CWJ acknowledges Institutional Support (Laboratory Directed Research and Development) at Los Alamos National Laboratory. The authors declare no competing interests. We thank Chris Marone for the laboratory data and Ke Gao for the numerical simulation data.

\section*{Author contributions statement}
KW developed the CED model and conducted the machine learning analysis with input from CWJ and PAJ. KW and KCB conceived of the transfer learning analysis and workflow. PAJ conduced experiments with collaborators at the Pennsylvania State University. PAJ was involved in collecting the simulation and experimental data. All authors were involved in writing the manuscript. 

\end{document}